\documentclass{article}
\PassOptionsToPackage{numbers}{natbib}
\usepackage[preprint]{neurips_2026}
\usepackage[utf8]{inputenc}
\usepackage[T1]{fontenc}
\usepackage{hyperref}
\usepackage{url}
\usepackage{booktabs}
\usepackage{amsfonts}
\usepackage{amsmath}
\usepackage{amssymb}
\usepackage{nicefrac}
\usepackage{microtype}
\usepackage{xcolor}
\usepackage{graphicx}
\usepackage{subcaption}
\usepackage{multirow}
\usepackage{algorithmic}
\usepackage{algorithm}
\usepackage{enumitem}
\usepackage{tabularx}
\usepackage{colortbl}
\usepackage{amsthm}
\usepackage{thmtools}
\usepackage{thm-restate}

\usepackage{tikz}
\usepackage{circledsteps}
\usepackage[section]{placeins}

\title{Phoenix TTS: High-Fidelity Synthesis and Voice Conversion via Flow-Matching-Driven Speech Tokenization}

\author{
    L-Lab  Phoenix-Audio Team\textsuperscript{1}
  \\
  \textsuperscript{1}Didichuxing Co. Ltd \\
}

\begin{document}

\maketitle

\begin{abstract}
  In current zero-shot text-to-speech (TTS) systems, conventional semantic tokenizers are typically optimized using supervised automatic speech recognition (ASR) or self-supervised learning (SSL) objectives. However, due to the inherent nature of speech, semantic and acoustic information cannot be completely decoupled, and ASR-based tokenizers discard acoustic details to focus on linguistic content; models relying on them usually struggle to achieve optimal speaker similarity. Furthermore, these tokenizers are optimized independently and lack direct supervision from downstream acoustic generation tasks. This isolated training creates a feature gap between the extracted discrete tokens and the continuous space required by acoustic models, fundamentally bottlenecking the upper bound of synthesis quality. To bridge this gap, we propose Phoenix TTS, a unified framework that tightly couples representation learning with generative acoustic modeling. Specifically, our speech tokenizer is optimized to reconstruct self-supervised features to maintain semantic richness, while simultaneously receiving direct supervision from a Flow Matching training loss. Through this joint training paradigm, the extracted discrete tokens successfully preserve essential semantic information and natively align with the feature space of the downstream Flow Matching model. Comprehensive evaluations highlight the efficiency and effectiveness of Phoenix TTS. Trained on 110K hours of data, the system achieves excellent speech intelligibility, yielding word error rates (WER) that consistently fall below those of ground-truth recordings. Simultaneously, it maintains robust zero-shot speaker similarity that rivals or outperforms several prominent large-scale baselines. Furthermore, as an advantageous byproduct of this unified training, the learned tokenizer can be seamlessly adapted to zero-shot voice conversion (VC) tasks without requiring task-specific fine-tuning.
\end{abstract}

\section{Introduction}

Modern zero-shot text-to-speech (TTS) is driven by two complementary objectives: linguistic fidelity, measured by Word Error Rate (WER), and speaker similarity, measured by how faithfully the synthesized voice matches a target prompt. To improve linguistic fidelity, recent systems primarily scale up training corpora, exploiting massive, diverse data to generalize across phonetic and prosodic contexts. To improve speaker similarity, the focus has shifted to architectural refinements and stronger speech representations that better capture and reproduce target acoustic characteristics.

Despite this progress, a critical bottleneck remains in the {discrete representation learning} stage. Prevailing speech tokenizers are derived from either Automatic Speech Recognition (ASR)~\citep{du2024cosyvoice,du2024cosyvoice1,du2025cosyvoice3} or Self-Supervised Learning (SSL)~\citep{wang2024maskgct,zhang2023speechtokenizer} feature reconstruction. ASR-based tokenizers~\citep{gao2023funasr,radford2023robust} were a major step forward for zero-shot TTS: their high compression ratio and semantically rich tokens are well suited to Large Language Model (LLM) modeling and substantially reduce the optimization burden on the generative side. However, semantic and acoustic information are not fully separable in speech, and tokenizers optimized purely for semantic accuracy inevitably discard fine-grained acoustic detail. As a result, TTS systems built on such tokenizers consistently achieve high intelligibility but fall short on speaker similarity. Compounding this, these tokenizers~\citep{zhou2026indextts2,anastassiou2024seed,du2024cosyvoice,du2025cosyvoice3,du2024cosyvoice1,wang2025spark,guo2024fireredtts} are trained entirely independently of the downstream acoustic model, without gradient-level supervision from the synthesis task. The resulting cascaded paradigm leaves a representation gap: discrete tokens are not natively aligned with the continuous acoustic space required for high-fidelity synthesis, capping both generative quality and zero-shot speaker adaptation.

To close this gap, we present Phoenix TTS, a unified framework that couples representation learning with generative acoustic modeling. Rather than freezing a pre-trained tokenizer, we train the speech tokenizer to reconstruct SSL features while simultaneously receiving gradient-level supervision from a Flow Matching loss imposed by the downstream decoder. Plugging the Flow Matching model directly into the tokenizer's reconstruction path turns it into the tokenizer's decoder, so that the resulting discrete tokens preserve semantic richness and are natively aligned with the continuous generation space. Despite being trained on a relatively modest 110K-hour corpus, Phoenix TTS achieves highly competitive results across rigorous zero-shot TTS benchmarks.

Our contributions are summarized as follows:
\begin{itemize}
    \item \textbf{Unified speech tokenizer framework.} We propose Phoenix TTS, in which the speech tokenizer is jointly optimized with a continuous Flow Matching decoder. By combining SSL feature reconstruction with downstream acoustic supervision, the framework preserves rich semantic information while natively aligning discrete tokens with the continuous generation space.
    
    \item \textbf{Native zero-shot voice conversion.} As a direct byproduct of joint training, the learned tokenizer supports high-fidelity zero-shot voice conversion (VC) without any task-specific fine-tuning, empirically validating its strong semantic--acoustic disentanglement.
    
    \item \textbf{Data efficiency and competitive performance.} Trained on only 110K hours of speech, Phoenix TTS attains a WER consistently lower than that of the ground-truth recordings while delivering competitive zero-shot speaker similarity, demonstrating favorable data efficiency.
\end{itemize}
\section{Related Work}
\subsection{Discrete Speech Representation}
Current discrete speech representations are generally categorized into acoustic tokens and semantic tokens. Acoustic tokens, typically extracted via neural speech codecs\cite{defossez2022highfi,xin2024bigcodec,kumar2023high,chen2025ds,zhang2023speechtokenizer}, are explicitly optimized for high-fidelity speech reconstruction by preserving fine-grained acoustic details. Conversely, semantic tokens are derived by quantizing continuous embeddings from SSL models\cite{chung2021w2v,hsu2021hubert} or by optimizing directly against ASR objectives. While this higher-level abstraction provides semantic tokens with exceptional robustness against acoustic noise and makes them inherently well-suited for discrete language modeling, it intrinsically sacrifices crucial acoustic details. 

\subsection{Zero-Shot TTS and Hybrid Architectures}
Zero-shot TTS aims to synthesize speech for unseen speakers by capturing timbre, prosody, and stylistic nuances from brief reference prompts. Empowered by in-context learning (ICL)\cite{chen2025neural}, recent frameworks have demonstrated remarkable improvements in naturalness and speaker adaptation. Current generative paradigms broadly fall into three categories: discrete codec language models, continuous diffusion or flow-matching models, and hybrid architectures.

While pure autoregressive (AR) approaches excel at in-context learning, they frequently suffer from error accumulation and unnatural prosodic breaks during long-form generation. Conversely, pure non-autoregressive (NAR) models\cite{chen2025f5,lee2024ditto,eskimez2024e2} generate high-fidelity audio but inherently struggle with long-form semantic alignment.

To bridge this divide, state-of-the-art hybrid systems employ a cascaded framework: an AR language model is utilized for text-to-semantic generation, paired with a diffusion-based or flow-based NAR decoder for semantic-to-acoustic synthesis. This hierarchical decoupling offloads the burden of capturing fine acoustic details to the continuous decoder, allowing the language model to focus exclusively on prosodic and linguistic alignment. Ultimately, this strategy systematically improves zero-shot generalization and ensures robust acoustic stability across unseen speakers.

Despite these architectural advancements, cascaded systems still rely on a static, independently trained speech tokenizer. Because the discrete semantic extraction process is completely severed from the downstream acoustic generation, an inherent feature mismatch persists as a critical bottleneck that our proposed unified training paradigm directly addresses.

\subsection{Continuous Representation TTS}
To mitigate the information loss inherent in discrete tokenization, recent studies explore continuous or semi-continuous speech representations \cite{meng2025autoregressive,liu2024autoregressive,chen2026sara}. Frameworks like DiTAR\cite{jia2025ditar} integrate autoregressive Language Models (LM) with continuous Diffusion Transformers (DiT)\cite{peebles2023scalable}. While achieving high acoustic fidelity, jointly optimizing a discrete autoregressive process with a continuous diffusion objective severely destabilizes training and impedes convergence.

  

\section{Phoenix TTS}
The proposed Phoenix TTS framework fundamentally comprises two primary components: the unified speech tokenizer (UniSpeechTokenizer) and an autoregressive LLM. Specifically, the encoder of the UniSpeechTokenizer is utilized to extract discrete semantic speech tokens. We subsequently employ the LLM to model the joint sequence of text and speech tokens, reformulating the TTS task as an autoregressive sequence generation problem. This generative process is conditioned on both text prompts and reference speaker prompts, with the latter being processed by a specialized learnable speaker encoder. Following the autoregressive generation, the Flow Matching Decoder within the UniSpeechTokenizer converts the predicted discrete tokens into continuous VAE latents via an optimal-path denoising process. Finally, a pre-trained Waveform VAE synthesizes high-fidelity, perceptible audio from these generated latents.

\subsection{UniSpeechTokenizer}
\label{unispeechtokenizer}
\begin{figure*}[htbp] 
  \centering
  
  
  
    


  \includegraphics[width=0.5\linewidth]{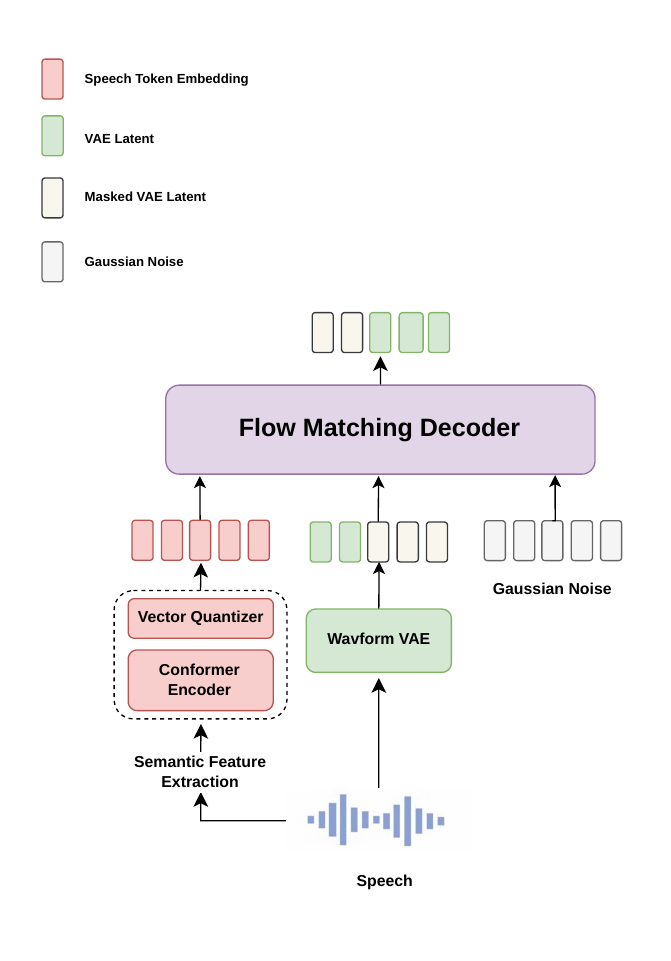}
  
  \caption{UnispeechTokenizer}
  \label{fig:tokenizer}
\end{figure*}
As illustrated in Figure \ref{fig:tokenizer}, we introduce the UniSpeechTokenizer, which derives discrete speech tokens by quantizing continuous features extracted from pre-trained speech SSL models. Because these SSL features exhibit profound correlations with linguistic and phonetic content, modeling them autoregressively is inherently more tractable than predicting raw acoustic tokens directly. Specifically, we extract the hidden states from the 16th layer of W2v-BERT 2.0\cite{chung2021w2v} to serve as continuous semantic representations, denoted as $\mathbf{x}$. These features are processed by a Conformer\cite{gulati2020conformer} encoder and discretized via a Vector Quantizer (VQ) with the codebook size of 8192, to yield the discrete speech token embeddings $\mathbf{e}_{sem}$:
\begin{align}
    \mathbf{e}_{sem}=\text{VQ}(\text{Conformer Encoder}(\mathbf{x}))\label{eq:vq}
\end{align}

To ensure that $\mathbf{e}_{sem}$ retains rich semantic information, it is passed through a Conformer decoder to reconstruct the original semantic features, yielding $\mathbf{\hat{x}}$. The semantic reconstruction objective is optimized via L1 loss:
\begin{align}
    \mathcal{L}_{feat}&=\left\|\mathbf{x}-\mathbf{\hat{x}}\right\|_1\label{eq:feat_loss}
\end{align}

Crucially, to effectively map these discrete semantic embeddings back into the high-fidelity acoustic generation space, we introduce a novel reconstruction paradigm utilizing a Flow Matching decoder. Parallel to the semantic extraction path, a pre-trained Waveform VAE, which follows from MegaTTS3 \cite{jiang2025megatts}, encodes the raw speech $s$ into continuous acoustic latents, denoted as $z_{vae} = \text{VAE\_Enc}(s)$. To circumvent the substantial optimization burden caused by frame-rate inconsistencies between semantic embeddings and VAE latents, we specifically select a VAE model whose downsampling factor perfectly matches the frame rate of the semantic tokens. This achieves a natural, one-to-one temporal alignment, entirely bypassing the complex optimization costs associated with resolving temporal resolution mismatches.

During training, we apply a binary mask $M$ to yield the masked representation $z_{mask} = M \odot z_{vae}$. We then formulate the overall condition vector $c$ by directly concatenating the discrete semantic tokens and the masked acoustic latents: $c = (e_{sem}, z_{mask})$.

To ensure the speech tokenizer remains highly adaptable to downstream acoustic generation, we jointly optimize the tokenizer alongside the Flow Matching decoder. Defining the target data distribution as $y_{1} = z_{vae}$ and the prior noise distribution as $y_{0} \sim \mathcal{N}(0, I)$, the Flow Matching model estimates the target vector field along the interpolation path $y_{t} = t y_{1} + (1 - t) y_{0}$ for continuous time steps $t \in [0, 1]$. The Flow Matching objective is defined as:

\begin{align}
\mathcal{L}_{FM}=\mathbb{E}_{t,\mathbf{y}_0,\mathbf{y}_1,\mathbf{c}}\left[\left\|v_\theta(\mathbf{y}_t,t,\mathbf{c})-(\mathbf{y}_1-\mathbf{y}_0)\right\|_2^2\right]
\end{align}
where $v_{\theta}$ is the predicted vector field parameterized by the Flow Matching decoder. Unlike conventional cascaded TTS systems, in which the tokenizer is trained in isolation and then frozen, our framework employs a holistic joint training paradigm. The total loss function $\mathcal{L}_{total}$ combines the flow matching loss, the quantization loss, and the semantic reconstruction loss:

\begin{align}
    \mathcal{L}_{total} = \lambda_{FM} \mathcal{L}_{FM} + \lambda_{VQ} \mathcal{L}_{VQ} + \lambda_{feat} \mathcal{L}_{feat}
\end{align}

Through this joint formulation, gradients derived from the acoustic reconstruction explicitly backpropagate into the Conformer encoder and the Vector Quantizer. This vital acoustic supervision forces the discrete tokens $e_{sem}$ to natively align with the Flow Matching generation space, robustly bridging the fundamental gap between semantic understanding and acoustic modeling.

\subsection{Autoregressive Language Modeling}
\begin{figure*}[htbp] 
  \centering
  
  
  
    


  \includegraphics[width=0.7\linewidth]{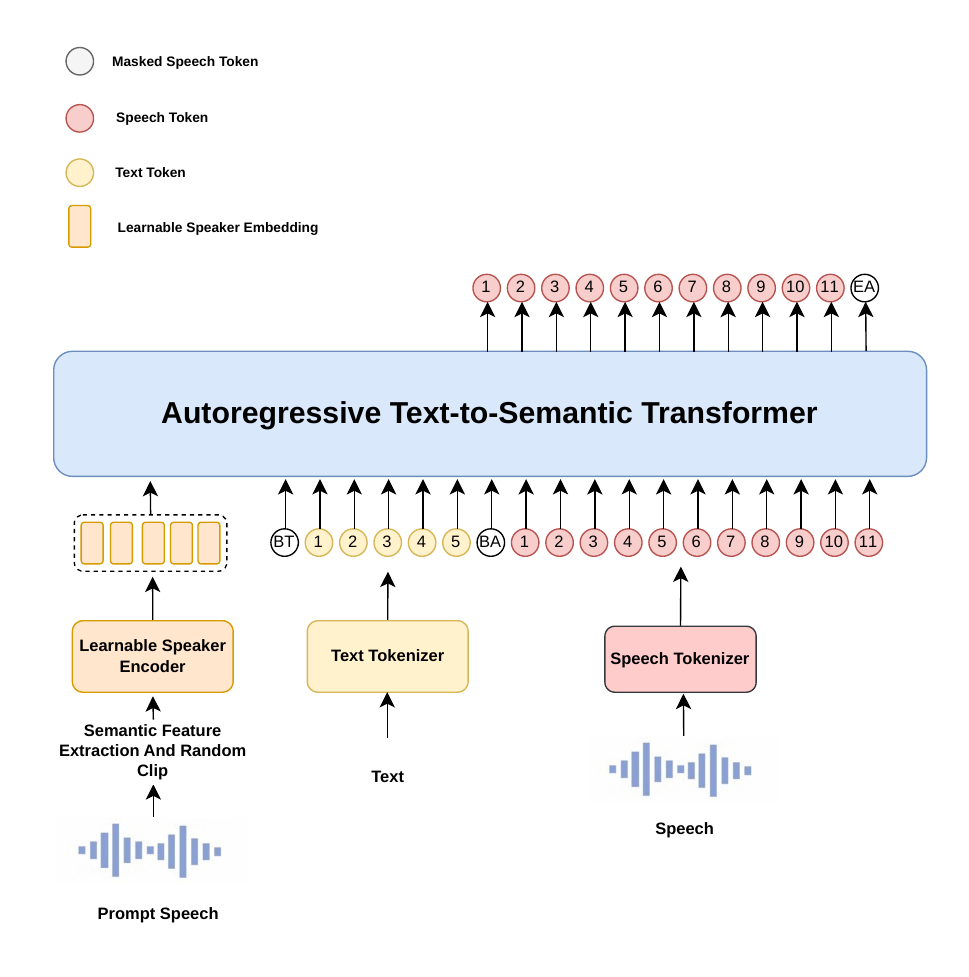}
  
  \caption{A schematic diagram of Phoenix TTS's autoregressive language modeling part. \Circled{BT}, \Circled{BA}, \Circled{EA} denote the ``start of text sequence'', ``start of speech'', and ``end of speech'' tokens.}
  \label{fig:ar_model}
\end{figure*}
As shown in Figure \ref{fig:ar_model}, we reframe the text-to-speech synthesis task as an autoregressive language modeling problem, built upon the pretrained Qwen2.5 0.5B architecture. The core objective of this module is to predict target discrete speech tokens conditioned on both the input text and a reference speaker prompt.

To enable zero-shot voice cloning, we first extract conditioning features from a reference prompt speech $y_{prompt}$. Following semantic feature extraction and random clipping, these features are mapped by a Learnable Speaker Encoder, which follows from IndexTTS2\cite{zhou2026indextts2} and XTTS\cite{casanova2024xtts} into a fixed-length sequence of continuous speaker embeddings, denoted as $E_{spk}$. Concurrently, the input text is processed by the Qwen2.5 0.5B Tokenizer into a sequence of discrete tokens $T = [t_{1}, t_{2}, \dots, t_{N}]$, while the target speech is encoded by our pre-trained UniSpeechTokenizer into discrete speech tokens $S = [s_{1}, s_{2}, \dots, s_{M}]$.

To construct the input sequence for the Autoregressive Text-to-Semantic LLM, we concatenate these representations alongside specific control tokens. Let $\langle \text{BT} \rangle$, $\langle \text{BA} \rangle$, and $\langle \text{EA} \rangle$ denote the special tokens for "Begin Text", "Begin Audio", and "End Audio", respectively. The conditional prefix $C$ is formulated as:
\begin{equation}
C = [E_{spk}, \langle \text{BT} \rangle, T, \langle \text{BA} \rangle]
\end{equation}

Given the prefix $C$, the Qwen2.5 0.5B model the joint probability of the target speech token sequence $S$ autoregressively. The probability of generating the entire speech sequence is factorized using the chain rule:
\begin{equation}
P(S|C) = \prod_{i=1}^{M} P_{\theta}(s_{i}|C, S_{<i})
\end{equation}
where $\theta$ represents the trainable parameters of the LLM, and $S_{<i} = [s_{1}, \dots, s_{i-1}]$ denotes the preceding speech tokens. The sequence terminates precisely when the model predicts the $\langle \text{EA} \rangle$ token.

During the training phase, the model is optimized using a standard teacher-forcing strategy. The optimization objective minimizes the negative log-likelihood (NLL) loss, effectively functioning as a cross-entropy loss over the speech token vocabulary:
\begin{equation}
\mathcal{L}_{AR} = -\mathbb{E}_{(C, S)} \left[ \sum_{i=1}^{M} \log P_{\theta}(s_{i}|C, S_{<i}) \right]
\end{equation}

By jointly attending to the continuous speaker embeddings and discrete text tokens, the LLM seamlessly captures both the target timbre and the linguistic content, successfully generating high-quality semantic tokens for the downstream Flow Matching decoder.

\section{Experiments}
\begin{table}[htbp]
\centering
\caption{Objective evaluation of Phoenix TTS and baseline models across different datasets, `-` denotes that the result is not available or not evaluated. *Metrics not reported in the original papers are calculated using the open-source checkpoints.}
\label{tab:objective_evaluation}
\resizebox{\textwidth}{!}{%
\begin{tabular}{l c c c c c c c c}
\toprule
\multirow{2}{*}{\textbf{Model}} & \multirow{2}{*}{\textbf{Training Data}} & \multicolumn{2}{c}{\textbf{LibriSpeech-PC}} & \multicolumn{2}{c}{\textbf{SeedTTS-EN}} & \multicolumn{2}{c}{\textbf{SeedTTS-ZH}} \\
\cmidrule(lr){3-4} \cmidrule(lr){5-6} \cmidrule(lr){7-8}
& & WER ($\downarrow$) & SIM ($\uparrow$) & WER ($\downarrow$) & SIM ($\uparrow$) & WER ($\downarrow$) & SIM ($\uparrow$) \\
\midrule
GT  & - &2.06 & 0.73 & 2.06 & 0.73 & 1.26 & 0.76 \\

\midrule
F5-TTS* & 100K Multi. & 2.42 & 0.660 & 1.83 & 0.670 & 1.56 & 0.760 \\
VoxCPM* & 1.8M Multi. & 2.22 & 0.710 & 1.58 & 0.729 & 1.38 & 0.773 \\
DiTAR &100K Multi. & 2.39 & 0.670 & 1.68 & \textbf{0.735} & \textbf{1.02} & 0.753 \\
\midrule

FireRedTTS & 170K Multi. &2.69 & 0.470 & 3.82 & 0.460 & 1.51 & 0.630 \\
MaskGCT  & 100K Multi. &2.63 & 0.687 & 2.62 & 0.717 & 2.27 & 0.774 \\
SparkTTS*  &100K Multi.& 2.81 & 0.575 &  3.14 & 0.573 & 1.54 & 0.660 \\
CosyVoice1* & 170K Multi. & 3.59 & 0.660 & 3.39 & 0.640 & 3.10 & 0.750 \\
CosyVoice2* & 200K Multi. &2.23 & 0.650 & 2.57 & 0.652 & 1.38 & 0.757 \\
CosyVoice3* & 530K Multi. &2.04 & 0.710 & 1.87 & 0.698 & 1.36 & 0.778 \\
IndexTTS2* & 55K Multi. &2.52 & 0.706 & 1.98 & 0.705 & 1.34 & 0.765 \\
\midrule
\textbf{Phoenix TTS} & 110K Multi. & \textbf{1.94} & \textbf{0.718} & \textbf{1.56} & 0.720 & 1.16 & \textbf{0.778} \\
\bottomrule
\end{tabular}%
}
\end{table}
\subsection{Experimental Setup}
\textbf{Training Datasets:} We trained our models on a comprehensive 110K-hour bilingual corpus, comprising 50K hours of Chinese and 60K hours of English speech data. This corpus was primarily curated from high-quality open-source datasets, specifically Emilia\cite{he2024emilia}, LibriHeavy\cite{kang2024libriheavy}, GigaSpeech\cite{chen2021gigaspeech}, and WenetSpeech4TTS\cite{ma2024wenetspeech4tts}, and strategically augmented with proprietary audiobook recordings to enhance domain diversity.
\\
\textbf{Training Details}
The UniSpeechTokenizer was optimized on 8 NVIDIA A100 (80GB) GPUs, utilizing a batch duration of 960 seconds per GPU over 10 epochs. Subsequently, the autoregressive language model (LLM) was trained across 32 NVIDIA A100 (80GB) GPUs, processing 400 seconds of audio per batch per GPU for 3 epochs. We employed the AdamW optimizer paired with a cosine learning rate scheduler, setting the peak learning rates to 1.0e-4 for the tokenizer and 3.0e-4 for the LLM.
\\
\textbf{LLM Training Data Preparation:} To facilitate robust zero-shot cloning, our LLM training dataset is structured at the speaker level. For each training instance, we sample two distinct utterances from the identical speaker: one serves as the speaker prompt, while the other acts as the generation target. Furthermore, we apply random cropping to the reference prompts during training to maximize data diversity and prevent model overfitting.
\\
\textbf{Architecture Configurations:}
We implement Phoenix TTS with a 0.5B-parameter configuration, comprising a 24-layer LLM initialized from the pre-trained Qwen2.5-0.5B, and a randomly initialized 18-layer Flow-Matching DiT for waveform-latent generation. The semantic tokenizer extracts 25Hz tokens from a frozen Wav2Vec2-BERT via a 6-layer Conformer encoder followed by a single-codebook VQ of size 8192. On the conditioning side, a 4-layer Conformer condenses prompt features, and a Perceiver Resampler with 32 latents projects them into the LLM embedding space. A pre-trained Waveform VAE compresses 24kHz audio into 25Hz, 64-dim latents, on which the DiT performs flow matching to synthesize the final waveform. Detailed configurations are shown in Table~\ref{tab:model_arch}.
\begin{table}[h]
\centering
\caption{The model architecture of PhoenixTTS.}
\label{tab:model_arch}
\begin{tabular}{@{}l p{0.62\linewidth}@{}}
\toprule
\textbf{Module} & \textbf{Configuration} \\
\midrule
Semantic Encoder      & Wav2Vec2-BERT (16kHz, layer-16) + 6-layer Conformer (dim 384), $\times 2$ downsample \\
Semantic VQ           & single codebook, size 8192, dim 8 (25Hz tokens) \\
Condition Encoder     & 4-layer Conformer, dim 1024$\rightarrow$512, FFN 2048, 4 heads, conv2d2 \\
Perceiver Resampler   & 32 latents, dim 896, ctx 512, 8 heads, FFN $\times 2$ \\
LLM                   & 24 layers (Qwen2.5-0.5B init.), dim 896, FFN 4864, 14 heads \\
Speech Vocabulary     & 8192 tokens + 2 special (BOS/EOS) \\
Flow-Matching DiT     & 18 layers, dim 768, 12 heads, FFN $\times 2$, semantic cond 1024 \\
Waveform VAE          & 24kHz $\rightarrow$ 25Hz latents (down [2,4,4,5,6]), latent dim 64 \\
Token Rate            & 25Hz (LLM and DiT both operate at 25Hz) \\
\bottomrule
\end{tabular}
\end{table}

\subsection{Evaluation}
To rigorously validate the generative capabilities of our framework, we evaluated Phoenix TTS across three standardized benchmarks: (1) SeedTTS test-en\cite{anastassiou2024seed}, comprising 1,000 utterances from the Common Voice dataset; (2) SeedTTS test-zh\cite{anastassiou2024seed}, containing 2,000 utterances sourced from DiDiSpeech\cite{guo2021didispeech}; and (3) LibriSpeech-PC-test-clean\cite{meister2023librispeech,panayotov2015librispeech}, featuring 1127 utterances.

\textbf{Objective Metrics}: We evaluate intelligibility via WER (using Whisper-Large-v3\cite{radford2023robust}\footnote{https://huggingface.co/openai/whisper-large-v3} for English and Paraformer\footnote{https://huggingface.co/funasr/paraformer-zh} for Chinese), timbre similarity via SIM (WavLM-TDNN\cite{chen2022wavlm}\footnote{https://github.com/microsoft/UniSpeech/tree/main/downstreams/speaker\_verification\#pre-trained-models}), and speech naturalness via UTMOS\footnote{https://github.com/sarulab-speech/UTMOS22}. 

\textbf{Subjective Metrics}: A crowd-sourced evaluation with 40 experienced native speakers was conducted to assess SMOS (speaker similarity, 1-5 scale) and CMOS (comparative quality against ground truth, -3 to 3 scale).

\textbf{Baseline Details} We evaluate our model against several state-of-the-art zero-shot TTS systems, including F5-TTS\footnote{https://huggingface.co/SWivid/F5-TTS}\cite{chen2025f5}, CosyVoice\footnote{https://huggingface.co/FunAudioLLM/CosyVoice-300M}\footnote{https://huggingface.co/FunAudioLLM/CosyVoice2-0.5B}\footnote{https://huggingface.co/FunAudioLLM/Fun-CosyVoice3-0.5B-2512}\cite{du2024cosyvoice,du2024cosyvoice1,du2025cosyvoice3}, IndexTTS2\footnote{https://huggingface.co/IndexTeam/IndexTTS-2}\cite{zhou2026indextts2}, SparkTTS\footnote{https://huggingface.co/SparkAudio/Spark-TTS-0.5B}\cite{wang2025spark}, MaskGCT\footnote{https://huggingface.co/amphion/MaskGCT}\cite{wang2024maskgct}, DiTAR\cite{jia2025ditar}, and VoxCPM\footnote{https://huggingface.co/openbmb/VoxCPM-0.5B}\cite{zhou2025voxcpm}.

\subsection{Zero-shot TTS Main Results}
Experimental results indicate that Phoenix TTS achieves exceptional performance in zero-shot speaker similarity across multiple languages. On the SeedTTS-ZH and LibriSpeech-PC test sets, our model attains SIM scores of 0.778 and 0.718, respectively, establishing a new state-of-the-art among all evaluated baselines (surpassing strong models like MaskGCT and IndexTTS2). Notably, in the Chinese synthesis scenario, our model's SIM score directly exceeds the objective metric of the Ground Truth reference (GT, 0.760). On the SeedTTS-EN set, Phoenix TTS remains highly competitive with a SIM of 0.720. These compelling results explicitly validate our core methodological hypothesis: by enabling explicit acoustic supervision from the Flow Matching decoder back to the frontend UniSpeechTokenizer, our framework successfully mitigates the loss of fine-grained acoustic nuances inherent in strictly decoupled training paradigms.

\begin{table}[htbp]

\centering
\caption{Subjective performance comparison on the zero-shot speech synthesis tasks on EN and ZH. SMOS scores are accompanied by 95\% confidence intervals (CI)}
\label{table:subjective}
\begin{tabular}{l c c c c}
\toprule
\multirow{2}{*}{\textbf{Model}}& \multicolumn{2}{c}{\textbf{EN}} & \multicolumn{2}{c}{\textbf{ZH}} \\
\cmidrule(lr){2-3} \cmidrule(lr){4-5}
& SMOS ($\uparrow$) & CMOS ($\uparrow$) & SMOS ($\uparrow$) & CMOS ($\uparrow$) \\
\midrule
GT & 4.22 $\pm$ 0.17 & 0 & 4.24 $\pm$ 0.15 & 0 \\
\midrule
Cosyvoice2 & 3.95 $\pm$ 0.20 & -0.31& 3.96 $\pm$ 0.20 & -0.22 \\
IndexTTS2 & 3.99 $\pm$ 0.18 & -0.11& 3.97 $\pm$ 0.20 & -0.12 \\
VoxCPM & 4.02 $\pm$ 0.17& -0.10& 4.03 $\pm$ 0.20 & -0.13 \\
\midrule
Phoenix TTS & \textbf{4.09 $\pm$ 0.16} &\textbf{ -0.09}& \textbf{4.10 $\pm$ 0.20} & \textbf{-0.10} \\
\bottomrule
\end{tabular}
\end{table}

While pushing the upper bounds of speaker similarity, Phoenix TTS concurrently achieves breakthrough performance in linguistic fidelity and pronunciation stability. On both the SeedTTS-EN (WER: 1.56) and LibriSpeech-PC (WER: 1.94) test sets, our model records the lowest error rates among all baselines, substantially outperforming the GT recordings (2.06 and 2.06, respectively). On the SeedTTS-ZH set, our model achieves a highly robust WER of 1.16, which also surpasses the GT performance (1.26) and outperforms the vast majority of architectures, trailing only DiTAR (1.02). These overall results demonstrate that our model does not compromise its underlying language understanding capabilities during the pursuit of acoustic fidelity. Benefiting from the strong linguistic priors of W2v-BERT 2.0 and our introduced semantic reconstruction loss ($\mathcal{L}_{feat}$), Phoenix TTS successfully preserves semantic integrity while natively aligning with the high-fidelity acoustic generation space.

\textbf{Subjective Evaluation} To complement the objective metrics, we conduct a subjective evaluation to assess the perceptual quality and speaker similarity of the generated speech in cross-sentence zero-shot synthesis tasks. We employ Similarity Mean Opinion Score (SMOS) to evaluate voice cloning fidelity and Comparative Mean Opinion Score (CMOS) to measure overall speech naturalness relative to the ground truth (GT). The evaluation encompasses both English (EN) and Chinese (ZH) test sets, with SMOS scores reported alongside 95\% confidence intervals.

As presented in Table \ref{table:subjective}, Phoenix TTS consistently outperforms all evaluated baselines across both metrics and languages. In terms of speaker similarity, our model achieves the highest SMOS of $4.09 \pm 0.16$ on the EN set and $4.10 \pm 0.20$ on the ZH set, surpassing strong baselines such as VoxCPM (4.02 and 4.03) and IndexTTS2 (3.99 and 3.97). This perceptual improvement aligns perfectly with our objective SIM results, verifying that the fine-grained acoustic details natively preserved by our joint modeling approach are distinctly recognizable by human listeners.

Furthermore, the CMOS results indicate that Phoenix TTS produces speech with naturalness that closely approaches human recordings. It achieves the highest CMOS scores among all evaluated models (-0.09 for EN and -0.10 for ZH), significantly narrowing the gap to the GT (0). In contrast, conventional cascaded systems or those relying on standard decoupled tokenizers, such as Cosyvoice2, exhibit a more pronounced degradation in comparative quality (-0.31 for EN and -0.22 for ZH).

Overall, these subjective evaluations firmly corroborate our objective findings. The human perceptual data demonstrate that our joint optimization framework achieves highly expressive and robust zero-shot generalization without sacrificing speech naturalness, establishing Phoenix TTS as a highly competitive architecture in the high-fidelity TTS domain.
\subsection{Speech Tokenizer Main Results}
\begin{table*}[htbp] 
\centering
\caption{Voice conversion performance on the LibriSpeech-PC test-clean set. We specifically select the Voice Conversion (VC) task to evaluate our UniSpeechTokenizer. Because the tokenizer is jointly trained with the Flow Matching decoder, the framework natively forms a highly effective VC pipeline, allowing us to directly assess the tokenizer's ability to preserve semantic integrity while adapting to unseen acoustic conditions. ‘w/o joint training’ demonstrates the speech tokenizer and the Flow Matching decoder are trained separately without joint acoustic supervision and w Mel Prompt demonstrates trained with traditional Mel-spectrogram with a mismatch frame rate.}
\label{tab:vc_performance}
\begin{tabular}{lccccccccc}
\toprule
\multirow{2}{*}{\textbf{Model}} & \multirow{2}{*}{\textbf{Frame Rate}} & \multirow{2}{*}{\textbf{Sample Rate}}  & \multicolumn{3}{c}{\textbf{Voice Conversion}} \\
\cmidrule(lr){4-6} 
& & &  WER ($\downarrow$)& SIM ($\uparrow$) & UTMOS ($\uparrow$)\\
\midrule
GT & - & - &  2.23 & 0.690 & 4.087 \\
Vocos & 93.75 & 24K  & 2.32  & 0.660 & 3.625\\
Vanilla VAE & 25 & 24K  & 2.41  & 0.683 & \textbf{4.095}\\
\midrule
\textbf{UniSpeechTokenizer} & 25 & 24K & 2.54  & \textbf{0.697} & 4.081 \\
\quad w/o joint training & 25 & 24K &2.70  & 0.682 & 4.036\\
\quad w Mel Prompt & 93.75 & 24K & 2.91  & 0.597 & 2.806\\
\bottomrule
\end{tabular}
\end{table*}

\begin{table*}[htbp] 
\centering
\caption{Zero-shot voice conversion results on Seed-TTS-Eval.}
\label{tab:vc_performance_seed}
\begin{tabular}{lccccccccc}
\toprule
\multirow{2}{*}{\textbf{Model}} & \multicolumn{2}{c}{\textbf{SeedTTS-EN}} & \multicolumn{2}{c}{\textbf{SeedTTS-ZH}} \\
\cmidrule(lr){2-3} \cmidrule(lr){4-5} 
& WER ($\downarrow$)& SIM ($\uparrow$)  &  WER ($\downarrow$)& SIM ($\uparrow$)\\
\midrule
GT & 1.96 & - & 1.33 & -\\
\midrule
Seed-VC & 2.57 & 0.56 & 2.52 &0.73\\
X-VC & 2.83 & 0.63  & 1.99& 0.73\\
\textbf{UniSpeechTokenizer} & \textbf{2.30} & \textbf{0.72}  &\textbf{1.98}  & \textbf{0.77} \\
\bottomrule
\end{tabular}
\end{table*}
To isolate and evaluate the representational capacity of the UniSpeechTokenizer, we assessed its performance on a VC task using the LibriSpeech-PC test-clean set. Because our tokenizer is jointly optimized with the Flow Matching decoder, it natively forms a highly effective zero-shot VC pipeline without any task-specific fine-tuning. This makes VC the optimal testbed for evaluating the tokenizer's disentanglement capabilities.

As detailed in Table \ref{tab:vc_performance}, despite operating at a highly compressed frame rate of 25 Hz, the UniSpeechTokenizer achieves a superior SIM score of 0.697, outperforming robust baselines like Vocos\cite{siuzdak2023vocos} (0.660) and Vanilla VAE (0.683). Remarkably, it marginally surpasses the Ground Truth (0.690), demonstrating a profound ability to aggregate target timbre while filtering out source-speaker artifacts. Concurrently, it maintains a highly competitive UTMOS of 4.081, confirming near-human naturalness.

Crucially, disabling the joint optimization mechanism (w/o joint training) results in a substantial performance degradation. In this traditional cascaded setting, the WER increases to 2.70, while SIM and UTMOS drop to 0.682 and 4.036, respectively. These results explicitly corroborate that isolating the tokenizer deprives it of essential gradient feedback from the acoustic decoder, leading to a loss of fine-grained acoustic information. Furthermore, replacing our frame-aligned VAE latents with traditional Mel-spectrogram prompts introduces a frame-rate discrepancy (93.75 Hz) and leads to a sharp decline in both SIM (0.597) and UTMOS (2.806). This collapse powerfully validates the effectiveness of our temporal alignment strategy, proving that matching the tokenizer's frame rate with the latent space is essential for high-fidelity synthesis.

To further validate the cross-lingual zero-shot voice conversion capabilities of our framework, we extended our evaluation to the Seed-TTS-Eval benchmark, encompassing both English (SeedTTS-EN) and Mandarin (SeedTTS-ZH) test sets. For this evaluation, we benchmarked our proposed UniSpeechTokenizer against two strong zero-shot VC baselines: Seed-VC\cite{liu2024zero} and X-VC\cite{zheng2026x}.

As illustrated in Table \ref{tab:vc_performance_seed}, our model demonstrates dominant performance across all metrics. Specifically, on the English subset, UniSpeechTokenizer achieves the highest speaker similarity (SIM: 0.72) and the lowest Word Error Rate (WER: 2.30), significantly outperforming X-VC (SIM: 0.63, WER: 2.83) and Seed-VC (SIM: 0.56, WER: 2.57). This superiority is equally evident in the Mandarin subset, where it establishes a peak SIM score of 0.77 while maintaining a highly robust WER of 1.98. These consistent results clearly indicate that the proposed tokenizer not only preserves accurate linguistic content—evidenced by low WERs approaching the Ground Truth—but also achieves exceptional semantic-acoustic disentanglement, enabling high-fidelity timbre transfer across different languages better than the baseline models.
\subsection{Ablation Study}
\begin{table}[htbp]
\centering
\caption{Ablation study evaluating the impact of speaker conditioning and joint optimization in Phoenix TTS. `w pretrained spk emb' refers to replacing the proposed Learnable Speaker Encoder with an explicit global pre-trained speaker embedding within the LLM. `w/o spk' indicates the absolute removal of the acoustic reference prompt from the LLM conditioning. Finally, `w/o joint training' demonstrates the baseline setting where the speech tokenizer and the Flow Matching decoder are trained separately without joint acoustic supervision.}

\label{tab:ablation}
\begin{tabular}{l c c c c}
\toprule
\multirow{2}{*}{\textbf{Model}}& \multicolumn{2}{c}{\textbf{SeedTTS-EN}} & \multicolumn{2}{c}{\textbf{SeedTTS-ZH}} \\
\cmidrule(lr){2-3} \cmidrule(lr){4-5}
& WER ($\downarrow$) & SIM ($\uparrow$) & WER ($\downarrow$) & SIM ($\uparrow$) \\
\midrule
GT & 2.06 & 0.73 & 1.26 & 0.76 \\
\midrule
Phoenix TTS & \textbf{1.56} & \textbf{0.720} & \textbf{1.16} & \textbf{0.778} \\
\quad w pretrained spk emb & 1.92 & 0.625 & 0.95 & 0.762 \\
\quad w/o spk & 2.20 & 0.422 & 2.70 & 0.560 \\
\quad w/o joint training & 2.36 & 0.706 & 1.33 & 0.764\\
\bottomrule
\end{tabular}
\end{table}

To validate the efficacy of our proposed architectural designs, specifically the speaker conditioning mechanism within the LLM and the joint training paradigm. We conduct a comprehensive ablation study on the SeedTTS-EN and SeedTTS-ZH datasets. The results, summarized in Table \ref{tab:ablation}, evaluate the impact of these components on speech intelligibility (WER) and zero-shot voice cloning fidelity (SIM).

As expected, the complete removal of the speaker reference prompt in the LLM (w/o spk) leads to a drastic degradation in zero-shot cloning capabilities. The SIM scores plummet from 0.720 to 0.422 on the English set, and from 0.778 to 0.560 on the Chinese set. Furthermore, the absence of a speaker prompt reference negatively impacts the autoregressive generation stability, resulting in a noticeable deterioration in WER (2.20 for EN and 2.70 for ZH). This confirms that explicitly conditioning the LLM with a speaker prompt is indispensable for guiding the initial semantic-acoustic alignment.

A critical comparison lies in substituting our Learnable Speaker Encoder with explicit, globally pre-trained speaker embeddings\footnote{https://github.com/alibaba-damo-academy/3D-Speaker/tree/main/egs/3dspeaker/sv-cam++} (w pretrained spk emb). Interestingly, while this global embedding variant yields a slight improvement in pronunciation robustness on the Chinese dataset (reducing WER from 1.16 to 0.95), it severely compromises speaker similarity across both languages, dropping the English SIM score substantially from 0.720 to 0.625. This phenomenon highlights a fundamental limitation of conventional TTS pipelines: compressing a reference utterance into a fixed-dimensional global embedding inevitably discards fine-grained, dynamic acoustic details. In contrast, our proposed Learnable Speaker Encoder directly preserves these subtle timbre and prosodic nuances, effectively raising the performance ceiling for high-fidelity voice cloning without relying on external speaker verification models.

Crucially, we evaluate the necessity of our holistic training approach by isolating the speech tokenizer and the Flow Matching decoder into a decoupled training pipeline (w/o joint training). This separation results in a severe degradation in text robustness, with the English WER spiking drastically from 1.56 to 2.36, alongside a consistent drop in SIM across both datasets. This empirical finding explicitly corroborates our core methodological hypothesis: optimizing the tokenizer in isolation severs the gradient feedback from the acoustic decoder, inevitably leading to a severe feature mismatch. Our joint training paradigm fundamentally bridges this gap, ensuring that the discrete semantic tokens are natively aligned with the high-fidelity acoustic generation space.

\section{Author}
\textbf{Core Contributors:}Peijie Chen, Zhuanling Zha, Zhipeng Nie, Weijie Wu, Yiming Liu, Daiyu Huang, Junbo Li, Jun Fang, Naiqiang Tan, Hua Chai\\
\textbf{Advisors:} Qingyang Hong(Xiamen University)
\section{Conclusion}
In this work, we address the inherent feature mismatch in mainstream zero-shot TTS systems caused by the strictly decoupled training of semantic tokenizers and downstream acoustic decoders. To bridge this fundamental gap, we introduced Phoenix TTS, a novel speech synthesis framework driven by a holistic joint training paradigm. Specifically, our architecture ensures semantic richness through the reconstruction of self-supervised learning (SSL) features, while simultaneously receiving explicit gradient-level supervision from a continuous Flow Matching decoder. This unified approach effectively resolves the information bottleneck of conventional tokenizers, ensuring that the learned discrete representations natively capture crucial acoustic nuances—such as timbre and prosody, without sacrificing linguistic integrity.

Extensive evaluations demonstrate the data efficiency and generative superiority of the proposed framework. Trained on merely 110K hours of data, Phoenix TTS achieves excellent speech intelligibility and significantly pushes the upper bound of zero-shot speaker similarity, consistently outperforming dominant large-scale baselines trained on substantially more data. Notably, in the Mandarin synthesis scenario, our model marginally exceeds the objective similarity metrics of ground-truth recordings. Furthermore, as an advantageous byproduct of this joint optimization, the learned tokenizer natively supports high-fidelity zero-shot voice conversion without requiring any task-specific fine-tuning. Ultimately, Phoenix TTS establishes a highly expressive, robust, and adaptable foundation for future high-fidelity speech generation tasks.


\bibliographystyle{plain} 
\bibliography{reference} 







\end{document}